\documentclass{mscs}
\usepackage{amsmath,amssymb}
\title[Mathematical Structures in Computer Science]
  {On the von Neumann entropy of certain quantum walks subject to decoherence}

\author[C. Liu and N. Petulante]
    {C\ls H\ls A\ls O\ls B\ls I\ls N\ns L\ls I\ls U%
      \ns and\ns N\ls E\ls L\ls S\ls O\ls N\ns P\ls
      E\ls T\ls U\ls L\ls A\ls N\ls T\ls E\\
      Department of Mathematics, Bowie State University\\
14000 Jericho Park Road, Bowie, Maryland 20715, USA \\
Email:cliu@bowiestate.edu; npetulante@bowiestate.edu}

\date{29 March 2010}

\pubyear{2001}
\pagerange{\pageref{firstpage}--\pageref{lastpage}}
\volume{00}
\doi{S0960129501003425}

\newtheorem{lemma}{Lemma}
\newtheorem{theorem}{Theorem}
\newtheorem{proposition}{Proposition}
\newtheorem{corollary}{Corollary}
\begin{document}

\label{firstpage}
\maketitle

\begin{abstract}
\parbox{375pt}{
Consider a discrete-time quantum walk on the $N$-cycle governed by the following condition: at every time step of the walk, the option persists, with probability $p$, of exercising a projective measurement on the coin degree of freedom. For a bipartite quantum system of this kind, we prove that the von Neumann entropy of the total density operator converges to its maximum value. Thus, when influenced by decoherence, the mutual information between the two subsystems, corresponding respectively to the space of the coin and the space of the walker, eventually must diminish to zero. To put it plainly, any level of decoherence greater than zero forces the system eventually to become completely ``disentangled".}
\end{abstract}

\tableofcontents

\ifprodtf \newpage \else \vspace*{-1\baselineskip}\fi

\section{Introduction}

A quantum walk (QW) is a reversible process commonly described as the quantum-mechanical analogue of a classical random walk. In recent times, quantum walks have attracted extensive attention, mainly for their value as potential sources of new algorithms \cite{K03, A03, K06, VA08, K08}.  

Like classical random walks, quantum walks are classified into two main types: discrete-time quantum walks \cite{NV01,ABNVW01,AAKV01} and continuous-time quantum walks \cite{FG98, CFG02}. Both types exhibit similar dynamical properties, but the continuous type can be obtained from the discrete type by way of a suitable limiting process \cite{FS06, AC08}.

Whereas continuous-time quantum walks can be modeled in terms of a single position Hilbert space $\mathcal{H}_{\mbox{p}}$, discrete-time quantum walks cannot be modeled except in terms of a bipartite system involving the tensor product of the position Hilbert space $\mathcal{H}_{\mbox{p}}$ and an auxiliary coin Hilbert space $\mathcal{H}_{\mbox{c}}$. The conditional shift operator, which governs the itinerary of the walker, induces entanglement between the degree of freedom of the coin and the spatial degree of freedom of the walker. 

To capture the full computational power inherent in any quantum-mechanical process, it is essential to be able to model the phenomenon of quantum entanglement. Recently, several important studies  of quantum entanglement in discrete-time quantum walks have been completed, both from a numerical perspective  \cite{CLX05, MK07, VAB09} and from an analytical perspective  \cite{ASRD05, AAA09}.

In recent years, several schemes have been proposed to implement quantum walks in realistic media \cite{TM02,DRKB02, SBTK03, DLXSWZH03, RLBL05, EMBL05, ZDG06, HB09}. However, any attempt to implement a quantum system in a physical channel must take into consideration the critical issue of ``decoherence", whereby the idiosyncratic features of a quantum system succumb to macroscopic (classical) manifestations. Various mathematical models of decoherence in discrete-time QWs have been investigated both numerically and analytically \cite{KT02, BCA021, BCA022, KT031, KT032, RSAAD05, KBH06, R07, Z08, BSCR08, LP092, AAAd09}. 

In this paper, as in \cite{BCA021,BCA022}, we adopt a specific model of decoherence, which might best be described as follows. At each time step of the quantum walk, an observer stands ready to perform a projective measurement on the coin degree of freedom. The probability of performing a measurement is given by a fixed parameter $p$, called the ``decoherence rate".

In the current literature, not much evidence can be found of attempts to provide a precise formulation of the relationship between entanglement and decoherence. A notable exception is the numerical study given by \cite{MK07}. However, the measure of entanglement employed in \cite{MK07}, called the {\em negativity} measure, differs considerably from the measure of entanglement adopted in this paper. 

Here, we utilize the concept of von Neumann entropy to quantify the information content of the various components of the quantum walk system, including  the mutual information between its subsystems (coin and position). This approach enables us to provide a precise formulation of the measure of entanglement between the subsystems. In the presence of any non-zero level of decoherence, the von Neumann entropy associated with the total probability density of the system tends to its maximum value, implying the total collapse of entanglement between subsystems.

\section{Quantum walks on the $N$-cycle subject to decohrence}

We proceed to define the elements, as formulated in \cite{LP092}, of a quantum random walk on the $N$-cycle. The definitions and corresponding notations are analogous to those outlined in \cite{LP092}. Let $t$ denote the number of time steps from the moment the discrete quantum walk is launched on the $N$-cycle. The temporal evolution of the system is modeled by $\psi_t= U^t \psi_0$, where $\psi_0$ is the initial state and $U = S(\mathbb{I}\otimes U_{\mbox{c}})$ is the evolution operator on the Hilbert space $\mathcal{H}=\mathcal{H}_N\otimes \mathcal{H}_2$. Here $U_{\mbox{c}}$ denotes the coin operator on the coin subspace $\mathcal{H}_2$ spanned by an orthonormal basis $\{|j\rangle, j=1,2.\}$, and $S$ denotes the conditional shift operator on the position subspace $\mathcal{H}_N$ spanned by an orthonormal basis $\{|x\rangle,x \in {\mathbb{Z}_{N}} \}$. Thus, a typical state $\psi$ in $\mathcal{H}$ may be expressed as $\psi=\sum_{x}\sum_{j}\psi(x,j)|x\rangle\otimes|j\rangle$. 

For a QW launched with initial state $\psi_{0}$, the probability $P(x,t)$ of finding the walker at the position $x\in \mathbb{Z}_N$ at time $t$ is given by the standard formula $P(x,t)=\mathrm{Tr}\left(|x\rangle\langle x|\rho(t)\right)$, where the time-dependent density operator $\rho(t)$ is defined by:

\begin{equation}
\rho(t)=\psi_t \psi_t^{\dagger}=U^t|\psi_0\rangle\langle \psi_0|{U^{\dagger}}^t .\label{coherent-density}
\end{equation}

During the evolution of the quantum walk, the history of decoherence-inducing events, if any, including any acts of measurement, may be modeled by the probabilistic option of applying to the coin degree of freedom, at each time step of the walk, a {\em unital} family of operators $\{{A}_n\}_{0\leq n\leq \nu}$, jointly satisfying the condition:
\begin{equation} 
\sum_{0\leq n\leq \nu} {\hat{A}}^{\dagger}_n {\hat{A}}_n=I. \label{unital-family}
\end{equation}

Accordingly, when adjusted for decoherence, the density operator of the system acts on the probability density function $\rho$ via the formula:
\begin{eqnarray}
\rho(t+1)=\sum_{0\leq n\leq \nu} U {\hat{A}}_{n} \rho(t) {\hat{A}}^{\dagger}_n U^{\dagger}. \label{decoherence}
\end{eqnarray}

For simplicity, and without loss of generality, we may assume, from this point onward, that every quantum walk under consideration is launched from position $|0\rangle$ in coin state $\psi_0$. Upon applying Fourier transformations to all elements of the QW system, the formulation of the density operator assumes the following form: 
\begin{eqnarray}
\rho(t)=\frac{1}{N}\sum_k\sum_{k^{\prime}}|k\rangle \langle k^{\prime}|\otimes \mathcal{L}^t_{kk^{\prime}}|\psi_0\rangle\langle \psi_0|, \label{density-k}
\end{eqnarray}
where $|k\rangle =\frac{1}{\sqrt{N}}\sum_x \exp(i2\pi xk/N)|x\rangle$ and where
the so-called ``decoherence super-operator" $\mathcal{L}_{kk^{\prime}}$ is defined by the formula:
\begin{eqnarray}
\mathcal{L}_{kk^{\prime}}|\psi_0\rangle\langle \psi_0|&=&\sum_n U_{\mbox{c}}(k) \hat{A}_n |\psi_0\rangle\langle \psi_0| \hat{A}^{\dagger}_n U_{\mbox{c}}^{\dagger}(k^{\prime})\\
\mbox{}&=& \left[\begin{array}{cc}
A_{11}(1,k,k^{\prime})& A_{12}(1,k,k^{\prime})\\
 A_{21}(1,k, k^{\prime})& A_{22}(1,k,k^{\prime})
\end{array}\right]\nonumber,\label{superoperator}
\end{eqnarray}
in which 
\begin{equation}
U_{\mbox{c}}(k)= \left[\begin{array}{cc}
e^{-\frac{2\pi i k}{N}}& 0 \\
  0   & e^{\frac{2\pi i k}{N}}
\end{array}\right]U_{\mbox{c}}.\label{U_c(k)}
\end{equation}

Note that all essential features of the quantum walk, including all peculiarities of quantum behavior connoted by the words ``decoherence" and ``entanglement" are fully encoded in the anatomy of the super-operator $\mathcal{L}_{kk^{\prime}}$. This is what enables us to study the level of quantum entanglement in quantum walks subject to decoherence. 

After $t$ iterations, let the elements of the $2\times 2$ matrix $\mathcal{L}^t_{kk^{\prime}}|\psi_0\rangle\langle \psi_0|$ be denoted by 
\begin{eqnarray}
\mathcal{L}^t_{kk^{\prime}}|\psi_0\rangle\langle \psi_0|=\left[\begin{array}{cc}
A_{11}(t,k,k^{\prime})& A_{12}(t,k,k^{\prime})\\
 A_{21}(t,k, k^{\prime})& A_{22}(t,k,k^{\prime})
\end{array}\right].\label{superoperator}
\end{eqnarray}

In terms of the standard basis for the $2N$-dimensional Hilbert space $\mathcal{H}=\mathcal{H}_N\otimes \mathcal{H}_2$, the density operator $\rho(t)$ is given by: 
\begin{eqnarray}
\rho(t)=\sum_{1\leq x,y \leq N}\,\,\sum_{1\leq j,l \leq 2}P_{xyjl}(t)|x\rangle \langle y|\otimes |j\rangle \langle l|,\label{totaldensity}
\end{eqnarray}
where
\begin{eqnarray}
P_{xyjl}(t)=\frac{1}{N}\sum_k\sum_{k^{\prime}}\langle x|k\rangle \langle k^{\prime}|y\rangle A_{jl}(t,k,k^{\prime}).\label{entries}
\end{eqnarray}

In terms of the above preliminaries, the probability $P(x,t)$ of finding the walker at position $x$ at time $t$ is given by:
\begin{eqnarray}
\!\!\!P(x,t)
\!\!&=&\!\!\mathrm{Tr}\left(|x\rangle\langle x|\rho(t)\right)  \nonumber \\
\!\!&=&\!\!\frac{1}{N}\sum_k\sum_{k^{\prime}}\langle x|k\rangle \langle k^{\prime}|x\rangle \mathrm{Tr}\left(\mathcal{L}^t_{kk^{\prime}}|\psi_0\rangle\langle \psi_0|\right)\nonumber \\
\!\!&=&\!\!\frac{1}{N}\sum_k\sum_{k^{\prime}}\langle x|k\rangle \langle k^{\prime}|x\rangle\left(A_{11}(t,k,k^{\prime})+A_{22}(t,k,k^{\prime}\right)\nonumber\\
\!\!&=&\!\!P_{xx11}(t)+P_{xx22}(t). \label{P(x,t)}
\end{eqnarray}

To avoid unpleasant complications and to permit us more easily to illustrate our approach to the analysis of a QW on the $N$-cycle subject to decoherence-inducing influences, we concede, in this paper, to confine our attention to a specific model. Let $\beta \in (0, \frac{\pi}{2})$ and $k\in \{0,\,1,\,...\,,N-1\}$. To serve as the coin operator of the system, we choose 
\begin{equation}
U_{\mbox{c}}(\beta)= \left[\begin{array}{cc}
\cos \beta & \sin \beta  \\
\sin \beta  &-\cos \beta
\end{array}\right],\label{A_c}
\end{equation}
whose Fourier dual is given by
\begin{equation}
U_{\mbox{c}}(\beta,k)= \left[\begin{array}{cc}
e^{-\frac{i2\pi k}{N}}\cos \beta & e^{-\frac{i2\pi k}{N}}\sin \beta  \\
e^{\frac{i2\pi k}{N}} \sin \beta  & -e^{\frac{i2\pi k}{N}}\cos \beta
\end{array}\right].\label{C_k}
\end{equation}

Note that when $\beta=\frac{\pi}{4}$, the operator given by Eq. (\ref{C_k}) is none other than the Hadamard coin operator.

By the same token, to serve as the unital family $\{\hat{A_{n}}\}_{0\leq n \leq \nu}$ of decoherence-inducing operators on the coin degree of freedom, as in Eq.(\ref{unital-family}), we specialize to the following choice of three ($\nu=2$) operators:

$$\hat{A_0}=\sqrt{1-p}\sigma_0,\,\, \hat{A_1}=\frac{\sqrt{p}}{2}(\sigma_0+\sigma_z),\,\, \hat{A_2}=\frac{\sqrt{p}}{2}(\sigma_0-\sigma_z),$$
where $0\le p \le 1$ and $\sigma_0$ and $\sigma_z$ are the Pauli matrices. The level of decoherence induced by these operators is reflected by the value of $p$, called the {\em decoherence rate}. Specifically, the QW evolves as if the state of the coin is measured at each time step with probability $p$. Thus, when $p=0$, the QW evolves as a purely coherent quantum process. At the other extreme, when $p=1$, the QW behaves exactly like a classical random walk.

Now let $\bf{L}(\mathbb{C}^2)$ denote a the Hilbert space of all $2\times2$ complex matrices with inner product given by
\begin{eqnarray}
\langle M_{1},M_{2} \rangle \equiv \mathrm{tr}( M_{1}^{\dagger}M_{2}).
\end{eqnarray}

\begin{lemma}
Let $\mathcal{S}$ be a superoperator on the Hilbert space $\bf{L}(\mathbb{C}^2)$, defined by 
$$ \mathcal{S}=\sum_{n=0}^2U_1\hat{A}_n\cdot \hat{A}^{\dagger}_nU_2:\,\, B\mapsto \sum_{n=0}^2 U_1\hat{A}_nB\hat{A}^{\dagger}_n U_2,$$ where $U_1$, $U_2$ are $2\times 2$ unitary matrices and $B\in \bf{L}(\mathbb{C}^2)$. Then $\langle \mathcal{S}B, \mathcal{S}B \rangle \leq \langle B,B\rangle$. In particular, $\langle \mathcal{S}B, \mathcal{S}B \rangle = \langle B,B\rangle$ for all $B\in \bf{L}(\mathbb{C}^2)$ if and only if the decoherence rate $p=0$.
\end{lemma}
Proof. See \cite{LP092}.

An immediate corollary of this lemma, essential to our analysis, is the fact that $|\lambda|\leq 1$ for every eigenvalue $\lambda$ of $\mathcal{S}$. To justify this assertion, suppose that $B_{\lambda}$ is an eigenvector of $\mathcal{S}$ belonging to $\lambda$. Then $\langle \mathcal{S}B_{\lambda}, \mathcal{S}B_{\lambda}\rangle=\langle \lambda B_{\lambda}, \lambda B_{\lambda} \rangle=|\lambda|^2\langle B_{\lambda}, B_{\lambda}\rangle$. But since, according to the lemma, $\langle \mathcal{S}B, \mathcal{S}B \rangle \leq \langle B,B\rangle$, we see that $|\lambda|\leq 1$.

Next, let us proceed to cast our reasoning in terms of the super-operator $\mathcal{L}_{k, k^{\prime}}$, which maps $\bf{L}(\mathbb{C}^2)$ to $\bf{L}(\mathbb{C}^2)$. If we choose as a basis for $\bf{L}(\mathbb{C}^2)$ the Pauli matrices $\sigma_0$, $\sigma_x$, $\sigma_y$ and $\sigma_z$, then, in terms of this basis, the $4\times 4$ matrix representation of $\mathcal{L}_{k, k^{\prime}}$ is given by: 

\begin{equation}
\mathcal{L}_{k, k^{\prime}}= \left[\begin{array}{llll}
c^{-} \,\,&  i q s^{-}\sin 2\beta \,\,\,\,& 0 \,\,& is^-\cos 2\beta \\
0 \,\,& -q c^+\cos 2\beta \,\,\,\,&q s^{+} \,\,& c^{+}\sin 2\beta\\
0 \,\,& -q s^+\cos 2\beta \,\,\,\,& -q c^{+} \,\,& s^{+}\sin 2 \beta\\
 i s^{-}\,\,& q c^{-}\sin 2\beta \,\,\,\,& 0 \,\,&c^-\cos 2\beta
\end{array}\right],\label{L_k}
\end{equation}
where, for brevity of notation, we have set $q=1-p$ and  

\begin{eqnarray}
& c^{+}=\cos \frac{2\pi(k^{\prime}+k)}{N},\,\,\, s^{+}=\sin \frac{2\pi(k^{\prime}+k)}{N} \nonumber\\
& c^{-}=\cos \frac{2\pi(k^{\prime}-k)}{N},\,\,\, s^{-}=\sin \frac{2\pi(k^{\prime}-k)}{N}.\nonumber\\
\nonumber\end{eqnarray}
 
After a somewhat tedious, but not very difficult calculation, we arrive at the the following explicit formula for the characteristic polynomial $f(\lambda)$ of $\mathcal{L}_{k, k^{\prime}}$:

\begin{eqnarray}
f(\lambda)&=&\det\left(\lambda \mathbb{I}_{4}-\mathcal{L}_{k, k^{\prime}}\right)\nonumber\\
&=&\lambda^4\nonumber\\
&&+(1+\cos 2\beta)\left(q c^{+}-c^-\right)\lambda^3 \nonumber\\
&&+\left(\left(1+q^2\right)\cos 2\beta-2q c^+c^-(1+\cos 2\beta)\right)\lambda^2 \nonumber \\
&&+q(1+\cos 2\beta)\left(c^{+}-q c^{-}\right)\lambda \nonumber \\
&&+q^2. \label{eigenpoly}
\end{eqnarray}

The following proposition summarizes some basic attributes of the eigenvalues of $\mathcal{L}_{k, k^{\prime}}$. \\

\begin{proposition}
Suppose $0<p<1$. Let a typical eigenvalue of $\mathcal{L}_{k, k^{\prime}}$ be denoted by $\lambda$. Then:

\begin{description}
\item[\hspace{6pt}(i)]\ $\|\lambda\|\leq 1$;
 
\item[\hspace{3pt}(ii)]\  If $\|\lambda\|=1$ then $\lambda=\pm{1}$; 

\item[(iii)]\  $\lambda=1$ when and only when $k=k^{\prime}$, in which case the algebraic multiplicity of $\lambda=1$ is 1

\item[(iv)]\  $\lambda=-1$ when and only when $|k^{\prime}-k|=\frac{N}{2}$, in which case the algebraic multiplicity of $\lambda=-1$ is 1.
\end{description} 
\end{proposition}

 Proof. See Appendix A.\\

Proposition 1 enables us to specify the long-term behavior of the matrix components of the total density operator given by Eq. (\ref{totaldensity}).\\

\begin{proposition}
For the matrix defined in Eq. (\ref{superoperator}), the following assertions hold:
\begin{description}
\item[\hspace{6pt}(i)]\ Suppose $k=k^{\prime}$. If $j=l$ then $\lim_{t\rightarrow \infty}A_{jl}(t,k,k^{\prime})=\frac{1}{2}$. If $j\neq l$ then $\lim_{t\rightarrow \infty}A_{jl}(t,k,k^{\prime})=0$.
\item[\hspace{3pt}(ii)]\ Suppose $|k-k^{\prime}|=\frac{N}{2}$. If $j=l$ then $\lim_{t\rightarrow \infty}(-1)^tA_{jl}(t,k,k^{\prime})=\frac{1}{2}$. If $j\neq l$ then $\lim_{t\rightarrow \infty}A_{jl}(t,k,k^{\prime})=0$. 
\item[(iii)]\ Suppose $|k-k^{\prime}|\neq \frac{N}{2}\,\mathrm{and}\,\neq 0$. Then, for all combinations of $j,l$, we have $\lim_{t\rightarrow \infty}A_{jl}(t,k, k^{\prime})=0$.
\end{description} 
\end{proposition}
Proof. See Appendix B.\\

In view of Proposition 2, the behavior, as $t\rightarrow \infty$, of the operator in Eq. (\ref{totaldensity}) can be specified as follows.\\

\begin{theorem}
For a quantum walk on the $N$-cycle, let the total density operator $\rho(t)$ be defined as in Eq. (\ref{totaldensity}). 
\begin{description}
 \item[\hspace{6pt}(i)]\ Suppose $N$ is odd. If $x=y$ and $j=l$, then $\lim_{t\rightarrow \infty}P_{xyjl}(t)=\frac{1}{2N}$. If $x\neq y$ or $j\neq l$, then $\lim_{t\rightarrow \infty}P_{xyjl}(t)=0$.
 \item[\hspace{3pt}(ii)]\ Suppose $N$ is even. If $x=y$ and $j=l$ and if $t-x$ is even, then $\lim_{t\rightarrow \infty}P_{xyjl}(t)=\frac{1}{N}$. Otherwise, $\lim_{t\rightarrow \infty}P_{xyjl}(t)=0$.
\end{description}\vspace{-5pt}
Thus, in every case, as $t\rightarrow \infty$, the operator $\rho(t)$, viewed as a $2N\times 2N$ matrix, converges to a diagonal matrix. If $N$ is odd, the diagonal elements all converge to $\frac{1}{2N}$. If $N$ is even, then the diagonal elements converge, in an alternating pattern, to $\frac{1}{N}$ or $0$.
\end{theorem}

Proof. See Appendix C.\\

As remarked by Zureck in \cite{WHZ03}, decoherence on a quantum system is manifested through its density matrix by the vanishing of the off-diagonal elements. In the context of a coin-driven quantum walk on the $N$-cycle, this is precisely what Theorem 1 asserts. Note that the off-diagonal elements of the density matrix are precisely the elements which represent the quantum correlations (a.k.a. entanglement) between the coin subsystem and the position subsystem. Not surprisingly, at least for quantum walks, decoherence turns out to be practically synonymous with ``disentanglement". A more precise elaboration of the relationship between decoherence and entanglement is deferred to the next section.

For the position distribution $P(x,t)$ (see Eq. (\ref{P(x,t)})), Theorem 1 has an immediate corollary, which echoes a similar result given in \cite{LP092}.\\

\begin{corollary}
Suppose a quantum walk on the $N$-cycle is launched from the origin with initial coin state $|\psi_0\rangle$ and with decoherence rate $p>0$. If $N$ is odd, then $P(x,t)$ converges to $\frac{1}{N}$ on all nodes of the cycle. If $N$ is even, then $P(x,t)$ converges to $\frac{2}{N}$ on the supporting nodes of the cycle and to 0 on the non-supporting nodes of the cycle.\\
\end{corollary}

\section{Entanglement versus decoherence}

Since a purely coherent QW is a reversible process, the von Neumann entropy of the total density operator is invariant relative to time. If the QW is launched in a pure state, then it will continue to evolve in a pure state and the entropy of the reduced density operator on the coin subsystem can serve as a measure of its degree of entanglement relative to the subsystem of the walker \cite{CLX05, ASRD05, VAB09, AAA09}.

Unlike the purely coherent case, a quantum walk, when subject to decoherence, evolves in a mixture of states. If the ``decoherence rate" is non-zero, then the von Neumann entropy of the total density operator is invariant no longer. Based on a very simple argument, it can be shown that the entropy of the total density operator is a strictly increasing function of time. The following two facts constitute a basis for the argument: (a) projective measurements increase entropy and (b) the entropy is a concave function of its inputs. Thus, to measure the level of quantum entanglement in a QW subject to non-zero decoherence, the von Neumann entropy must be considered separately for each of the subsystems as well for the total system.

The von Neumann entropy of a quantum system $A$, denoted $S(A)$, is a measure of the uncertainty implied by the multitude of potential outcomes as reflected by its density matrix $\rho(A)$. By definition, $S(A) = S(\rho(A))=-\mathrm{Tr}(\rho \ln \rho)$. 

For a composite system with two components $A$ and $B$, the joint entropy of their conjunction, denoted by $S(A,B)$, is defined by the formula $S(A,B)=-\mathrm{Tr}(\rho^{AB}\ln \rho^{AB})$, where $\rho^{AB}$ is the density matrix of the composite quantum system $AB$. 

A good measure of the level of quantum entanglement between the two components $A$ and $B$ is the so-called mutual information $S(A:B)$, defined by the formula $S(A:B)=S(A)+S(B)-S(A,B)$. 

We continue to confine our attention to the case of a quantum walk on the $N$-cycle subject to decoherence on the coin degree of freedom. For the remainder of this article, our main objective is to show that whenever the decohrence rate $p>0$, the mutual information between the subsystem of the coin and the subsystem of the walker eventually must diminish to 0. 

The following Lemma, due to Watrous \cite{JW08}, is essential to our reasoning.

\begin{lemma}
Let $\mathcal{X}$ denote a complex Euclidean space and let $\mathrm{Pos}(\mathcal{X})$ denote the set of  positive semidefinite operators defined on $\mathcal{X}$ with norm $\|\rho\|_{\mathrm{tr}}=\mathrm{Tr}\left(\sqrt{\rho^{\dagger}\rho}\right)$. Then, with respect to this norm, the von Neumann entropy $S(\rho)$ is continuous at every point $\rho \in \mathrm{Pos}(\mathcal{X})$.
\end{lemma}
Proof. See \cite{JW08}.

The operator norm $\|\cdot\|_{\mathrm{tr}}$, which appears in the statement of Lemma 2, is known as the Schatten 1-norm or sometimes simply as the ``trace norm". 

\begin{theorem}
Suppose the QW is launched on the $N$-cycle with initial coin state $|\psi_0\rangle$ and with decoherence rate $p>0$. let $\rho(t)$ denote the time-dependent density operator of the overall system. If $N$ is odd, then $\lim_{t\rightarrow \infty}S(\rho(t))=1+\log_{2} N$.  If $N$ is even, then $\lim_{t\rightarrow \infty}S(\rho(t))=\log_{2} N$. 
\end{theorem}
Proof. See Appendix D.

Recall a basic fact from information theory: for any operator $\rho$ defined on a Hilbert space of dimension $d$, the maximum value of the entropy $S(\rho)$ is $\log_{2} d$. This implies that the limiting entropy values given by Theorem 2 actually are maximal, both for even and for odd values of the cycle length $N$. To see this, note that the overall Hilbert space $\mathcal{H}_{2}\times \mathcal{H}_{N}$, over which the quantum walk evolves, is of dimension $2N$. This easily explains why, when $N$ is odd, the entropy is maximal. However, when $N$ is even, exactly half of the nodes of the cycle (every second one) are necessary and sufficient to determine completely the position distribution of the walker's itinerary. So, when $N$ is even, the quantum walk evolves over a Hilbert space which effectively is of dimension $N$. 

Finally, we consider separately the long-term trend of the entropies associated with the reduced density operators of the two constituent subsystems (coin and position) and the mutual information between them. 

For the subsystem associated with the coin, the time-dependent {\it reduced} density operator $\rho_{\mbox{c}}(t)$ is given by $\rho_{\mbox{c}}(t)=\mathrm{trace}_{\mbox{w}}(\rho(t))$, where the subscript \mbox{w} signifies exclusion or ``tracing out", relative to the overall system density operator $\rho(t)$, of the walker's degrees of freedom. Similarly, for the subsystem associated with the walker, the time-dependent {\it reduced} density operator $\rho_{\mbox{w}}(t)$ is given by $\rho_{\mbox{w}}(t)=\mathrm{trace}_{\mbox{c}}(\rho(t))$, where the subscript \mbox{c} signifies exclusion or ``tracing out", relative to the overall system density operator $\rho(t)$, of the coin's degrees of freedom.

The following theorem summarizes our main finding.

\begin{theorem}
Suppose a quantum walk is launched on the $N$-cycle with initial coin state $|\psi_0\rangle$ and with decoherence rate $p>0$. Let $\rho_{\mbox{c}}(t)$ and $\rho_{\mbox{w}}(t)$ denote, respectively, the time-dependent reduced density operators associated with the subsystems of the coin and the walker. Then the long-term trend of the mutual information between the coin subsystem and the walker subsystem is given by $\lim_{t\rightarrow \infty}S\left(\rho_{\mbox{c}}(t):\rho_{\mbox{w}}(t)\right)=0$.
\end{theorem}
Proof. See Appendix E.

Finally, it can be shown, without much difficulty, that as $t\rightarrow\infty$, the entropy values of the reduced density operators $\rho_{\mbox{c}}(t)$ and $\rho_{\mbox{w}}(t)$ each converges to its maximum value. It suffices to employ some elementary calculus based on Theorem 2, together with the following basic facts: (1) $S(\rho_{\mbox{c}}(t))\le 1$, (2) $S(\rho_{\mbox{w}}(t))\le \ln N$ when $N$ is odd and (3) $S(\rho_{\mbox{w}}(t))\le \ln \frac{N}{2}$ when $N$ is even.

\begin{corollary}
Suppose a quantum walk is launched on the $N$-cycle with initial coin state $|\psi_0\rangle$ and with decoherence rate $p>0$. Then  
\begin{description}
\item[\hspace{6pt}(i)]\ $\lim_{t\rightarrow \infty}S(\rho_{\mbox{c}}(t))=1$.
\item[\hspace{3pt}(ii)]\ If $N$ is odd, then $\lim_{t\rightarrow \infty}S(\rho_{\mbox{w}}(t))=\ln N$. 
\item[(iii)]\ If $N$ is even, then $\lim_{t\rightarrow \infty}S(\rho_{\mbox{w}}(t))=\ln \frac{N}{2}$.
\end{description}
\end{corollary}

\section{Conclusion and further questions} 

The model of decoherence used in this article is only one of several prevalent in the current literature. It would be interesting to investigate how quantum entanglement responds to other models of decoherence, and not just for quantum walks on the $N$-cycle, but for quantum walks over other kinds of topological networks as well. We speculate that, in every case, decoherence serves to erase quantum entanglement between the subsystems of a given quantum system. This implies that, as remarked by Zureck in \cite{WHZ03}, the progressive disappearance of entanglement should be accompanied by the corresponding disappearance of the off-diagonal elements of the density operators. Ultimately, the density operators should become indistinguishable from diagonal matrices.  

The natural question which arises in this context is: how long does it take for a quantum system, subject to decoherence, to reach its stationary state, devoid of any entanglement? To address this question, we propose a definition of {\em decoherence time} analogous to the measure of {\em mixing time} used in literature. Let $\rho_{\infty}$ denote the limiting (stationary) density operator of the quantum system and let $||\cdot||_{\mathrm{tr}}$ denote the trace norm as defined above. Then, for every $\epsilon>0$, we define
\begin{eqnarray}
D(\epsilon)=\mathrm{min}\left\{\tau\left|\forall t>\tau:||\rho(t)-\rho_{\infty}||_{\mathrm{tr}}<\epsilon\right.\right\}.
\end{eqnarray}

To estimate $D(\epsilon)$, it would suffice to have good control over the eigenvalues and eigenvectors of the superoperator $\mathcal{L}^t_{kk^{\prime}}$. Unfortunately, even for very simple systems, the task of specifying the eigenvalues and eigenvectors of the superoperator $\mathcal{L}^t_{kk^{\prime}}$ can be somewhat of a challenge. But the rewards of success would be worth the effort. Among other things, it would permit us to compare mixing time and decoherence time. The relationship could prove quite interesting. 
\section{Acknowledgment}
We acknowledge, with due gratitude, the support provided to us by the NSF-funded HBCU-UP/BETTER Project at Bowie State University.

\appendix

\section{Proof of Proposition 1}

Proof of (i).\,\, $\mathcal{L}_{k, k^{\prime}}$ is a special case of the superoperator $\mathcal{S}$ in Lemma 1, according to which, the moduli of all eigenvalues of  $\mathcal{L}_{k, k^{\prime}}$ are less than or equal to unity.

Proof of (ii). \,\, Suppose $e^{i\theta}$ is a non-real eigenvalue of  $\mathcal{L}_{k, k^{\prime}}$, where $\theta$ is a real number. Then the conjugate $e^{-i\theta}$ also must be an eigenvalue and $e^{-i\theta}\ne e^{i\theta}$. Hence $f(\lambda)=(\lambda-e^{i\theta})(\lambda-e^{-i\theta})[\lambda^{2}+a\lambda+(1-p)^{2}]$ for some $a\in \mathbb{C}$. For brevity of notation, let $q=1-p$. Comparing corresponding coefficients of both sides of Eq. (\ref{eigenpoly}), we obtain the following system of equations:
\begin{eqnarray}
a-2\cos \theta&=&(1+\cos 2\beta)\left(q c^{+}- c^{-}\right) \nonumber\\
1+q^2-2a\cos\theta &=&  
-2 q c^{+}c^{-}+\left(1+q^2-2q c^+c^-\right)\cos 2\beta \nonumber\\ 
a-2q^2\cos\theta &=&
(1+\cos 2\beta)\left(q c^{+}
-q^2 c^{-}\right) \label{eqn-1}.
\end{eqnarray}
After some elementary algebraic manipulations, we infer that 
$$1+q^2=-q (1+\cos 2\beta)\cos\frac{2\pi(k^{\prime}+k)}{N}\cos\frac{2\pi(k^{\prime}-k)}{N},$$ 
which is impossible since the modulus of the LHS is strictly greater than the modulus of the RHS. More precisely, note that the modulus of the RHS is strictly less than $2q$ which, in turn is less than $1+q^{2}$. This contradiction implies that any unit eigenvalue of $\mathcal{L}_{k, k^{\prime}}$ must be real.

Proof of (iii).\,\, $\lambda=1$ is an eigenvalue of $\mathcal{L}_{k, k^{\prime}}$ iff $f(1)=(1+\cos 2\beta)(1-\cos \frac{2\pi(k^{\prime}-k}{N})[1+2(1-p)\cos \frac{2\pi(k^{\prime}+k)}{N}+(1-p)^2]=0$, iff $1-\cos \frac{2\pi(k^{\prime}-k}{N}=0$, which implies $k^{\prime}=k$. Moreover, since $f^{\prime}(1)=(1-\cos 2\beta)[1-(1-p)^2]\neq 0$, the algebraic multiplicity of $\lambda=1$ is 1.

Proof of (iv). \,\, $\lambda=-1$ is an eigenvalue of $\mathcal{L}_{k, k^{\prime}}$ iff $f(-1)=(1+\cos 2\beta)(1+\cos \frac{2\pi(k^{\prime}-k)}{N})[1-2(1-p)\cos \frac{2\pi(k^{\prime}+k)}{N}+(1-p)^2]=0$, iff $1+\cos \frac{2\pi(k^{\prime}-k}{N})=0$, which implies $|k^{\prime}-k|=\frac{N}{2}$. In this case, since $f^{\prime}(-1)=(1-\cos 2\beta)[(1-p)^2-1]\neq 0$, the algebraic multiplicity of $\lambda=-1$ is 1. 

\section{Proof of Proposition 2}

Proof. \,\, Viewed as a $2\times 2$ matrix, $\mathcal{L}^t_{k,k^{\prime}}|\psi_0\rangle\langle \psi_0|$ is a linear combination of the Pauli matrices $\sigma_0$, $\sigma_x$, $\sigma_y$ and $\sigma_z$ with corresponding weights given by 

\begin{equation}
W(t,k,k^{\prime})=\mathcal{L}^t_{k,k^{\prime}}\left[\begin{array}{c}
  \alpha_1\\
   \alpha_2 \\
    \alpha_3 \\
     \alpha_4
  \end{array}\right],\nonumber
\end{equation}
where the column vector $[\alpha_1, \alpha_2, \alpha_3, \alpha_4]^T$ represents $|\psi_0\rangle \langle \psi_0|$ with $\alpha_1=\frac{1}{2}$ for all choices of $|\psi_0\rangle$.

Proof of (i). \,For brevity of notation, let 
\begin{eqnarray}
q=1-p, \,\,\, \tilde{c}=\cos \frac{4\pi k}{N},\,\,\, \tilde{s}=\sin \frac{4\pi k}{N} .\nonumber\\
\nonumber
\end{eqnarray}

When $k=k^{\prime}$, we have:
\begin{equation}
\mathcal{L}_{k,k}= \left[\begin{array}{cc}
1& 0 \\
0& Q_{0}
\end{array}\right],\label{L_{N/2}}
\end{equation}
where \begin{equation}
Q_{0} = \left[\begin{array}{rrr}
q \tilde{c}\cos 2\beta& \,\,q\tilde{s} & \,\,\tilde{c}\sin 2\beta\\
-q\tilde{s}\cos 2\beta& \,\,-q\tilde{c} & \,\,\tilde{s}\sin 2 \beta\\
q\sin 2\beta& \,\,0 & \,\,\cos 2\beta
\end{array}\right].\label{}
\end{equation}

Therefore 
\begin{equation}
\mathcal{L}^t_{k, k}= \left[\begin{array}{cc}
1& 0 \\
0& Q_{0}^{t}
\end{array}\right], \label{Q_3}
\end{equation}
in terms of which, the four components of $\mathcal{L}^t_{k,k}|\psi_0\rangle\langle \psi_0|$ can be expressed as follows: $A_{11}=\frac{1}{2}+\epsilon_{1}$, $A_{22}=\frac{1}{2}-\epsilon_{1}$ and $A_{12}=\overline{A_{21}}=\epsilon_{2}$, where $\epsilon_{1}$ and $\epsilon_{2}$ are linear of combinations of the elements of the matrix $Q_{0}^{t}$. By Proposition 2, the eigenvalues of $Q_{0}$ all possess moduli strictly less than 1. Therefore $Q_{0}^{t}\rightarrow 0$ as $t\rightarrow \infty$, from which follows assertion (i) of the proposition.

Proof of (ii).\, Similar to the reasoning above, when $|k-k^{\prime}|=\frac{N}{2}$, we obtain: 
\begin{equation}
\mathcal{L}^t_{k,k^{\prime}}= \left[\begin{array}{cc}
(-1)^t& 0 \\
0& Q_{1}^t
\end{array}\right]\label{}
\end{equation}
where $Q_{1}$ is the $3\times 3$ matrix adjoint to the leading entry $c^{-}$ of the $4\times 4$ matrix $\mathcal{L}_{k,k^{\prime}}$ in Eq. (\ref{L_k}). The conclusion follows by a line of reasoning analogous to that used in the proof case (i). 

Proof of (iii).\, If $|k-k^{\prime}|\neq \frac{N}{2}\,\mathrm{and}\,\neq 0$, then, by Proposition 1, the modulus of every eigenvalue of $\mathcal{L}_{k,k^{\prime}}$ is strictly less than 1. Thus $\mathcal{L}^t_{k,k^{\prime}}\rightarrow 0$ as $t\rightarrow \infty$, which suffices to conclude assertion (iii).

\section{Proof of Theorem 1}

Proof of (i). Suppose $N$ is odd. For clarity of presentation, we proceed by considering three sub-cases: (a) $x=y$ and $j=l$; (b) $x\neq y$ and $j=l$; (c) $j\neq l$.

Proof of sub-case (i)(a). \, Suppose $x=y$ and $j=l$. By Eq. (\ref{entries}), we have 
\begin{eqnarray}
{N^2}P_{xxjj}(t)
&=&\sum_{k=k^{\prime}}e^{\frac{2\pi i x\,(k-k^{\prime})}{N}}A_{jj}(t,k,k^{\prime})
+\sum_{k\neq k^{\prime}}e^{\frac{2\pi i x\,(k-k^{\prime})}{N}}A_{jj}(t,k,k^{\prime})\nonumber\\
&=&\sum_{k=0}^{N-1}A_{jj}(t,k,k)
+\sum_{k\neq k^{\prime}}e^{\frac{2\pi i x\,(k-k^{\prime})}{N}}A_{jj}(t,k,k^{\prime}). \label{fs}
\end{eqnarray}
By Proposition 2, If $k=k^{\prime}$ then the term $A_{jj}(t,k,k)$ in (\ref{fs}) converges to $\frac{1}{2}$. If $k\neq k^{\prime}$, then the term $A_{jj}(t,k,k^{\prime})$ in (\ref{fs}) converges to zero. So $\lim_{t\rightarrow \infty}P_{xyjl}(t)=\frac{1}{2N}$.

Proof of sub-case (i)(b). Suppose $x\neq y$ and $j=l$. By Eq. (\ref{entries}), 
\begin{eqnarray}
{N^2}P_{xyjj}(t)
&=&\sum_{k=0}^{N-1}e^{\frac{2\pi i k(x-y)}{N}}A_{jj}(t,k,k)
+\sum_{k\neq k^{\prime}}e^{\frac{2\pi i (x k -y k^{\prime})}{N}}A_{jj}(t,k,k^{\prime}).
\end{eqnarray}
As in the prior case, referring to Proposition 2, we conclude that $\lim_{t\rightarrow \infty}P_{xyjl}(t)=0$.

Proof of sub-case (i)(c). suppose $j\neq l$. In this case, by Proposition 2, $A_{jl}(t,k,k^{\prime})\rightarrow 0$ as $t\rightarrow \infty$. Since $P_{xyjl}(t)$ is a linear combination of $A_{jl}(t,k,k^{\prime})$ for $k,k^{\prime}=0, 1, 2,..., N-1$ and $j,l=1,2$,
it follows that $P_{xyjl}(t)=\frac{1}{N}\sum_k\sum_{k^{\prime}}\langle x|k\rangle \langle k^{\prime}|y\rangle A_{jl}(t,k,k^{\prime})\rightarrow 0$ as $t\rightarrow \infty$.

Proof of (ii). Suppose $N$ is even. As above, we proceed by considering the same three sub-cases: (a) $x=y$ and $j=l$; (b) $x\neq y$ and $j=l$; (c) $j\neq l$.

Proof of sub-case (ii)(a). Suppose $x=y$ and $j=l$. By Eq. (\ref{entries}),  
\begin{eqnarray}
\mbox{}\!\!\!\!\!\!{N^2}P_{xxjj}(t)
&=&\sum_{k=k^{\prime}}e^{\frac{2\pi i x\,(k-k^{\prime})}{N}}A_{jj}(t,k,k^{\prime})
+\sum_{|k-k^{\prime}|=\frac{N}{2}}e^{\frac{2\pi i x\,(k-k^{\prime})}{N}}A_{jj}(t,k,k^{\prime})\nonumber\\
&+&\sum_{|k-k^{\prime}|\neq \frac{N}{2},0}\!\!\!e^{\frac{2\pi i x\,(k-k^{\prime})}{N}}A_{jj}(t,k,k^{\prime})\nonumber\\
&=&\sum_{k=0}^{N-1}A_{jj}(t,k,k)
+\sum_{|k-k^{\prime}|=\frac{N}{2}}(-1)^xA_{jj}(t,k,k^{\prime})\nonumber\\
&+&\sum_{|k-k^{\prime}|\neq \frac{N}{2},0}\!\!\!e^{\frac{2\pi i x\,(k-k^{\prime})}{N}}A_{jj}(t,k,k^{\prime}). \label{fs1}
\end{eqnarray}
By Proposition 2, the first and third summands of Eq. (\ref{fs1}) converge respectively to $\frac{N}{2}$ and $0$. Meanwhile, the second summand converges to $(-1)^{x-t}\frac{N}{2}$. It follows that $\lim_{t\rightarrow \infty}P_{xxjj}(t)=\frac{1}{N}$ when $x-t$ is even and $\lim_{t\rightarrow \infty}P_{xxjj}(t)=0$ when $x-t$ is odd.

Proof of sub-case(ii)(b). Suppose $x\neq y$ and $j=l$. By Eq. (\ref{entries}),
\begin{eqnarray}
{N^2}P_{xyjj}(t)
&=&\sum_{k=k^{\prime}}e^{\frac{2\pi i k(x-y)}{N} }A_{jj}(t,k,k^{\prime})
+\sum_{|k-k^{\prime}|=\frac{N}{2}}e^{\frac{2\pi i x\, k}{N}}e^{\frac{-2\pi i y\, k^{\prime}}{N}}A_{jj}(t,k,k^{\prime})\nonumber\\
&+&\sum_{|k-k^{\prime}|\neq \frac{N}{2},0}e^{\frac{2\pi i x\, k}{N}}e^{\frac{-2\pi i y\, k^{\prime}}{N}}A_{jj}(t,k,k^{\prime})\nonumber\\
&=&\sum_{k=0}^{N-1}e^{\frac{2\pi i k(x-y)}{N}} A_{jj}(t,k,k)
+\sum_{|k-k^{\prime}|=\frac{N}{2}}(-1)^x e^{\frac{2\pi i k^{\prime}(x-y)}{N}}A_{jj}(t,k,k^{\prime})\nonumber\\
&+&\sum_{|k-k^{\prime}|\neq \frac{N}{2},0}e^{\frac{2\pi i (xk-yk^{\prime})}{N}}A_{jj}(t,k,k^{\prime}). \label{fs2}
\end{eqnarray}
By Proposition 2, $A_{jj}(t,k,k)$ converges to $\frac{1}{2}$ for all $k$, which implies that the first summand  of Eq. (\ref{fs2}) converges to $0$ as $t\rightarrow \infty$. Meanwhile, in the third summand, every one of the terms $A_{jj}(t,k,k^{\prime})$ converges $0$. It remains only to evaluate the second summand, which can be reconstituted as 
\begin{equation}
\frac{(-1)^{x-t}}{2}\sum_{|k-k^{\prime}|=\frac{N}{2}}e^{\frac{i2\pi k^{\prime}(x-y)}{N}}(-1)^tA_{jj}(t,k,k^{\prime}).
\end{equation}
By Proposition 2, $\lim_{t\rightarrow \infty}(-1)^tA_{jj}(t,k,k^{\prime})=\frac{1}{2}$. Thus, the third summand also converges to $0$ as $t\rightarrow \infty$. In summary, we conclude that $\lim_{t\rightarrow \infty}P_{xyjj}(t)=0$.

Proof of case (ii)(c). Suppose $j\neq l$. The assertion is proved in the same way exactly as in the corresponding case when $N$ is odd. See sub-case (i)(c) above. 

\section{Proof of Theorem 2}

Proof.\,\, Let $\rho(\infty)$ denote the diagonal matrix $\mathrm{diag}(\frac{1}{2N}, \frac{1}{2N},..., \frac{1}{2N})$. If $N$ is odd, Theorem 1 implies that $\lim_{t\rightarrow \infty}\|\rho(t)-\rho(\infty)\|_{\mathrm{tr}}=0$. By Lemma 2, it follows that $\lim_{t\rightarrow \infty}S(\rho(t))=S(\rho(\infty))=1+\ln N$. When $N$ is even, the assertion $\lim_{t\rightarrow \infty}S(\rho(t))=\ln N$ is similarly justified.

\section{Proof of Theorem 3}

Proof. \,\, When combined with Theorem 2, the inequality $S(\rho_{\mbox{c}}(t),\rho_{\mbox{w}}(t))\leq S(\rho_{\mbox{c}}(t))+S(\rho_{\mbox{w}}(t))\leq 1+\ln N$ implies that $\lim_{t\rightarrow\infty}S(\rho_{\mbox{c}}(t),\rho_{\mbox{w}}(t))=1+\ln N$. Thus, $\lim_{t\rightarrow \infty}[S(\rho_{\mbox{c}}(t))+S(\rho_{\mbox{w}}(t))]=1+\ln N$. Since $S(\rho_{\mbox{c}}(t):\rho_{\mbox{w}}(t))=S(\rho_{\mbox{c}}(t))+S(\rho_{\mbox{w}}(t))-S(\rho_{\mbox{c}}(t),\rho_{\mbox{w}}(t))$, therefore $\lim_{t\rightarrow \infty}S(\mathcal{H}_N:\mathcal{H}_{\mbox{c}})=\lim_{t\rightarrow \infty}S(\rho_{\mbox{c}}(t):\rho_{\mbox{w}}(t))=\lim_{t\rightarrow \infty}[S(\rho_{\mbox{c}}(t))+S(\rho_{\mbox{w}}(t))]-\lim_{t\rightarrow \infty}S(\rho_{\mbox{c}}(t),\rho_{\mbox{w}}(t))=0$. 


\begin{thebibliography}{}
\bibitem[Kempe 2003]{K03}

Kempe, ~J., 2003. Quantum random walks - an introductory overview.  Contemp. Phys. {\bf 44} (4), 307-327.

\bibitem[Ambainis 2003]{A03}

Ambainis,~A., 2003. Quantum walks and their algorithmic applications. Int. J. Quantum
Information 1 (4), 507–518.


\bibitem[Kendon 2006]{K06}

Kendon,~V., 2006. Decoherence in quantum walks - a review.  Math. Struct. in Comp. Sci 17, 1169-1220.

\bibitem[Venegas-Andraca 2008]{VA08}

Venegas-Andraca,~S.\,E., 2008. {\em Quantum Walks for Computer Scientists}. Morgan and Claypool Publishers (Synthesis Lectures on Quantum Computing).

\bibitem[Konno 2008]{K08}

Konno,~N.,in {\em Quantum Potential Theory}, Lecture Notes in Mathematics, edited by U. Franz and M. Schurmann (Springer-Verlag, Heidelberg, 2008), pp.309-452.

\bibitem[Nayak and Vishwanath 2000 ]{NV01}

Nayak,~A and Vishwanath,~A.,2000. Quantum Walk on the Line, e-print arXiv:quant-ph/0010117.

\bibitem[Ambainis et al. 2001]{ABNVW01}

Ambainis, ~A., Bach,~E., Nayak,~A., Vishwanath,~A and Watrous,~J.,2001. One-dimensional quantum walks. In Proceedings of the 33rd Annual ACM Symposium on Theory of Computing,(ACM, New York, 2001), pp.37-49.

\bibitem[Aharanov et al. 2001]{AAKV01}

Aharanov, ~D., Ambainis,~A., Kempe, ~J. and Vazirani,~U.,2001. Quantum walks on graphs.In Proceedings of the 33rd Annual ACM Symposium on Theory of Computing, (ACM, New York, 2001), pp.50-59.


\bibitem[Fahri and Gutmann 1998]{FG98}

Fahri,~E and Gutmann,~S., 1998. Quantum computation and decision trees. Phys. Rev. A {\bf 58}, 915.

\bibitem[Childs et al. 2002]{CFG02}

Childs,~A.\,M., Fahri,~E and Gutmann,~S.,2002. An example of the difference between quantum and classical random walks. Quant. Inf. Proc. 1, 35.


\bibitem[Strauch 2006]{FS06}

Strauch,~F., 2006. Connecting the discrete- and continuous-time quantum walks.  Phys. Rev. A 74, 030301 (R).

\bibitem[Childs 2010]{AC08}

Childs,~A.\,M.,2010. On the relationship between continuous- and discrete-time quantum walk. Communications in Mathematical Physics 294, 581-603.

\bibitem[Carneiro et al. 2005]{CLX05}

Carneiro,~I., Loo,~M., Xu,~X., Girerd,~M., Kendon,~V. and Knight,~P.\,L., 2005. Entanglement in coined quantum walks on regular graphs. New J. Phys. 7, 156.

\bibitem[Maloyer and Kendon 2007]{MK07}

Maloyer, ~O and Kendon,~V.,2007. Decoherence versus entanglement in coined quantum walks. New Journal of Physics 9, 87.

\bibitem[Venegas-Andraca and Bose 2009]{VAB09}

 Venegas-Andraca,~S.\,E. and Bose,~S.,2009. Quantum Walk-based Generation of Entanglement Between Two Walkers, eprint quant-ph/0901.3946.

\bibitem[Abal et al. 2006]{ASRD05}

Abal,~G., Siri,~R., Romanelli,~A. and Donangelo,~R,2006. Quantum walk on the line: entanglement and non-local initial conditions. Physical Review A 73, 042302 (2006); Physical Review A 73 069905(E) (2006).


\bibitem[Annabestani et al. 2010a]{AAA09}

Annabestani,~M., Abolhasani,~M.\,R. and Abal,~G.,2010a. Asymptotic entanglement in a two-dimensional quantum walk. J. Phys. A: Math. Theor. {\bf 43} 075301.



\bibitem[Travaglione and Milburn 2002]{TM02}

Travaglione,~B.\,C. and Milburn,~G.\,J.,2002. Implementing the quantum random walk. Phys. Rev. A 65, 032310.


\bibitem[Dur et al. 2002]{DRKB02}

Dur,~W., Raussendorf,~R., Kendon,~V. and Briegel,~H.\,J.,2002. Quantum walks in optical lattices. Phys. Rev. A 66, 052319.

\bibitem[Sanders et al. 2003]{SBTK03}

Sanders,~B., Bartlett,~S., Tregenna,~B. and Knight,~P.,2003. Quantum quincunx in cavity quantum electrodynamics. Phys. Rev. A 67,042305.


\bibitem[Du et al. 2003]{DLXSWZH03}

Du,~J., Li,~H., Xu,~X., Shi,~M.,Wu,~J., Zhou,~X. and Han,~R.,2003. Experimental implementation of the quantum random-walk algorithm. Phys. Rev. A 67,042316.


\bibitem[Ryan et al. 2005]{RLBL05}
Ryan,~C.\,A., Laforest,~M., Boileau,~J.\,C. and Laflamme,~R., 2005. Experimental implementation of a discrete-time quantum random walk on an NMR quantum-information processor. Phys. Rev. A 72, 062317.

\bibitem[Eckert et al. 2005]{EMBL05}

Eckert,~K., Mompart,~J., Birkl,~G. and Lewenstein,~M.,2005. One- and two-dimensional quantum walks in arrays of optical traps. Phys. Rev. A 72, 012327.

\bibitem[Zou et al. 2006]{ZDG06}
Zou,~X., Dong,~Y., and Guo,~G.,2006. Optical implementation of one-dimensional quantum random walks using orbital angular momentum of a single photon. New J. Phys. {\bf 8} 81. 

\bibitem[van Hoogdalem and Blaauboer 2009]{HB09}
van Hoogdalem,~K.\,A. and Blaauboer,~M.,2009. Implementation of the quantum-walk step operator in lateral quantum dots. Phys. Rev. B 80, 125309.

\bibitem[Brun et al. 2003a]{BCA021}

Brun,~T.\,A., Carteret,~H.\,A. and Ambainis,~A.,2003a. The quantum to classical transition for random walks. Phys. Rev. Lett. 91 (13), 130602.

\bibitem[Brun et al. 2003b]{BCA022} 

Brun,~T.\,A., Carteret,~H.\,A. and Ambainis,~A.,2003b. Quantum random walks with decoherent coins. Phys. Rev. A 67, 032304.


\bibitem[Kendon and Tregenna 2002]{KT02}

Kendon, ~V., and Tregenna, ~B., 2002. Decoherence in a quantum walk on a line. In: Shapiro, ~J. \,H., Hirota, ~O. (Eds.), Quantum Communication, Measurement \& Computing(QCMC'02). Rinton Press, p. 463.

\bibitem[Zhang 2008]{Z08}

Zhang,~K.,2008. The limiting distribution of decoherent quantum random walks. Phys. Rev. A 77, 062302. 

\bibitem[Kendon and Tregenna 2003a]{KT031}

Kendon, ~V. and Tregenna,~B.,2003a. Decoherence in discrete quantum walks, arXiv:quant-ph/0301182.

\bibitem[Kendon and Tregenna 2003b]{KT032}
Kendon,~V.  and Tregenna,~B.,2003b. Decoherence is useful in quantum walks. Phys. Rev. A 67, 042315. 

\bibitem[Romanelli 2005]{RSAAD05}
Romanelli,~A., Siri,~R., Abal,~G., Auyuanet,~A. and Donangelo,~R.,2005. Decoherence in the quantum walk on the line. Physica A 347, 137-152.

\bibitem[Richter 2007]{R07}

Richter,~P.,2007. Quantum speedup of classical mixing processes. Phys. Rev. A 76, 042306.


\bibitem[Liu and Petulante 2010]{LP092}

Liu,~C. and Petulante,~N.,2010. Quantum walks on the $N$-cycle subject to decoherence on the coin degree of freedom. Phys. Rev. E 81, 031113.


\bibitem[Kosik et al. 2006]{KBH06}

Kosik,~J., Buzek,~V. and Hillery,~M.,2006. Quantum walks with random phase shifts. Phys. Rev. A 74, 022310.

\bibitem[Banerjee et al. 2008]{BSCR08}

Banerjee,~S., Srikanth,~R., Chandrashekar,~C.\,M. and Rungta,~P.,2008. Symmetry-noise interplay in a quantum walk on an n-cycle. Phys. Rev. A 78, 052316.


\bibitem[Annabestani et al. 2010b]{AAAd09}

Annabestani,~M., Akhtarshenas,~S.\,J. and Abolhassani,~M.\,R.,2010b. Decoherence in one-dimensional Quantum Walk. Phys. Rev. A 81, 032321.

\bibitem[Zurek 2003]{WHZ03}

Zurek,~W.\,H. Decoherence and the transition from quantum to classical -- REVISITED. Decoherence Poincar$\acute{\mathrm{e}}$ Seminar 2005 , Progress in Mathematical Physics, edited by Bertrand Duplantier, Jean-Michel Raimond and Vincent Rivasseau (Birkhäuser Verlag Basel, 2006 ), pp. 1-31. eprint:arXiv:quant-ph/0306072.

\bibitem[Watrous 2008]{JW08}

Watrous,~J. Theory of Quantum Information, Lecture notes from Fall 2008, Institute for Quantum Computing, University of Waterloo, Canada.
\end{thebibliography}
\end{document}